\documentclass[aps,prb,a4paper,showpacs,preprint]{revtex4-1}

\usepackage{amsmath}

\newcommand{\Gbold}{\bf G\rm}
\newcommand{\Ibold}{\bf I\rm}
\newcommand{\Sigmabold}{\boldsymbol{\Sigma}}
\newcommand{\sigmabold}{\boldsymbol{\sigma}}

\newcommand{\Phibold}{\boldsymbol{\Phi}}
\newcommand{\ivec}{\boldsymbol{i}}
\newcommand{\jvec}{\boldsymbol{j}}
\newcommand{\nvec}{\boldsymbol{n}}
\newcommand{\mvec}{\boldsymbol{m}}
\newcommand{\kvec}{\boldsymbol{k}}
\newcommand{\qvec}{\boldsymbol{q}}

\newcommand{\avec}{\boldsymbol{a}}
\newcommand{\Kvec}{\boldsymbol{K}}
\newcommand{\Qvec}{\boldsymbol{Q}}
\newcommand{\Rvec}{\boldsymbol{R}}
\newcommand{\zerovec}{\boldsymbol{0}}
\newcommand{\deltavec}{\boldsymbol{\delta}}

\usepackage{graphicx}

\begin{document}

\title{Enhancement of gaps in thin graphitic films for heterostructure formation}

\author{J.P. Hague}
\affiliation{The Open University, Walton Hall, Milton Keynes, MK7 6AA, United Kingdom}

\begin{abstract}

There are a large number of atomically thin graphitic films with
similar structure to graphene. These films have a spread of bandgaps
relating to their ionicity, and also to the substrate on which they
are grown. Such films could have a range of applications in digital
electronics where graphene is difficult to use. I use the dynamical
cluster approximation to show how electron-phonon coupling between
film and substrate can enhance these gaps in a way that depends on the
range and strength of the coupling. One of the driving factors in this
effect is the proximity to a charge density wave instability for
electrons on a honeycomb lattice. The enhancement at intermediate
coupling is sufficiently large that spatially varying substrates and
superstrates could be used to create heterostructures in thin
graphitic films with position dependent electron-phonon coupling and
gaps, leading to advanced electronic components.
\end{abstract}

\date{25th February 2013}


\maketitle

\section{Introduction}

The 2D material graphene has made headlines over the past decade for
its remarkable properties. Often overlooked is the availability of
other two-dimensional graphitic materials. These graphitic (graphite
like) materials are not formed from carbon atoms, but have a similar
structure and properties to graphene, but with a direct bandgap that
is lacking in suspended graphene. These gapped compounds have the
potential to make graphene compatible digital transistors,
semiconductor lasers and solar cells, and it would be impossible to
make such devices without a band gap. The hope is that 2D graphitic
compounds with a bandgap have both the exotic properties of materials
such as graphene, with the major technological importance of 3D
semiconductors.

Atomically thick graphitic materials with honeycomb lattices and an
inherent direct bandgap formed because of strong ionicity include
boron nitride (BN) \cite{novoselov2005a} (band gap $\sim 5.6$eV
\cite{song2010a}) and other materials can be grown in very similar
hexagonal Wurtzite layers, such as InN (band gap 0.7-0.8eV)
\cite{wu2002a}, InSb (0.2eV \cite{khan1968a}, possibly down to 45meV
on certain substrates \cite{bedi1997a}), GaN ( 2.15eV
\cite{petalas1995a}), and AlN (6.28eV \cite{litimein2002a}), again due
to inherent ionicity. It has also been reported that small gaps due to
local ionicity can be formed with a similar mechanism in
graphene-gold-ruthenium systems \cite{enderlein2010a} and graphene-SiC
systems (there is some debate about the latter
\cite{zhou2007a,bostwick2007a}). Finally, the 2D layered materials
MoSe$_2$ \cite{tongay2012a} and MoS$_2$ \cite{novoselov2005a} also
have useful gaps and properties, although they are not considered here
as the honeycomb like structure has three atoms per unit cell: two Se
or S atoms for each Mo atom.


Recently, I used a self-consistent mean-field theory to show that gaps
in atomically thin materials with a honeycomb structure may be
modified by introducing strong electron-phonon coupling through a
highly polarizable superstrate \cite{hague2011b,hague2012a}. Similar
interactions between graphitic monolayers and substrates form
polaronic states and affect the overall electronic structure of the
monolayers, as shown by quantum Monte Carlo simulations for highly
doped thin graphitic films \cite{hague2012b}. Strong effective
electron-electron interactions can be induced via coupling between the
electrons in atomically thick monolayer and phonons in a highly
polarizable substrate because of limited out of plane screening,
similar to that seen for quasi-2D materials such as cuprates where the
dimensionless electron-phonon coupling can be of order unity
\cite{alexandrov2002a}. 
Dimensionless electron-phonon couplings of up to $\lambda=1$ have been
reported in systems of graphene on various substrates from angle
resolved photoemission spectroscopy studies (see Fig. 3 of
Ref. \onlinecite{siegel2012a} and references therein\footnote{There
  are two electron-phonon couplings in graphene, one between electrons
  in the plane and phonons in the plane, and another between electrons
  in the plane and phonons in the substrate. Coupling between
  electrons and in-plane phonons vanishes at half filling, as is the
  case for graphene on metals where weak coupling is expected with the
  substrate, whereas the electron-phonon interaction measured for
  graphene on SiC has no significant doping dependence, indicative
  that the coupling is with the substrate. In most cases, the coupling
  measured with ARPES is several times higher than would be expected
  if there were no coupling to the substrate.}), and large couplings are found
in intercalated graphite compounds, including a measured
$\lambda=0.45$ in KC$_8$ \cite{gruneis2009a}. Since the experimental
trend in graphene has been to keep the electron-phonon coupling as
small as possible so that record mobilities can be obtained in
graphene sheets, a coordinated effort in the other direction could in
principle lead to very large couplings that cause novel features in
the band structure.

Previous theoretical work on the electron-phonon interaction in
graphene has focussed on monolayer graphene without
ionicity. Signatures of electron-phonon coupling with substrates can
be found in ARPES spectra \cite{calandra2007a,tse2007a}. In suspended
or decoupled graphene monolayers, properties such as the Fermi
velocity are not significantly renormalized by electron-phonon
coupling \cite{park2007a,tse2007a} (there is insufficient space to
review all studies of electron-phonon interaction in the various forms
of graphene, but a review of the earlier work in this area, including
the effects on transport can be found in
Ref. \onlinecite{castroneto2009a}). The work here differs because it
studies graphitic materials such as thin films of III-V semiconductors
where ionicity is present (represented as a static potential that
differs for A and B sites). I make calculations beyond the mean-field
theory by using the dynamical cluster approximation formalism (DCA) to
compute the effects of electron-phonon interaction on electrons in
atomically thick graphitic materials, where a gap has been opened
because of ionicity. I present results computed with a high order
iterated perturbation theory consistent with Migdal's theorem (which
allows neglect of vertex corrections for low phonon frequency and weak
coupling) and discuss the effect of long range interactions.

Besides the use of electron-phonon interactions with substrates, the
possibility of tunable gaps has mainly focused on graphene. Following
a theoretical proposal \cite{mccann2006a,mccann2007a}, bilayer
graphene has been observed to have a gap that can be tuned by applying
an external electric field \cite{ohta2006a,zhang2009a}. Electron
confinement in graphene nanoribbons leads to gaps \cite{han2007a}, and
high quality nanoribbons can be made by unzipping nanotubes
\cite{kosynkin2009a} or using patterned SiC steps
\cite{hicks2012a}. Very wide bandgaps can be formed by functionalizing
graphene with hydrogen (graphane)
\cite{sofo2007a,boukhvalov2008a,elias2009a} and fluorine
(fluorographene) \cite{charlier1993a,cheng2010a}. 

This paper is organized as follows: A model Hamiltonian for the
interactions between graphitic monolayers and substrates is introduced
in Sec. \ref{sec:model}. The perturbative expansion and dynamical
cluster formalism used to solve the model are discussed in
Sec. \ref{sec:method}. Sec. \ref{sec:results} presents details of gap
enhancements and the spontaneous formation of a charge density wave
state. A summary and conclusions are presented in
Sec. \ref{sec:conclusions}.

\section{Model Hamiltonian}
\label{sec:model}

The Hamiltonian required to describe the motion of electrons in thin films with
honeycomb lattices has a basis of two atoms. Typically, electron motion within the
plane is described using a tight binding model, and ionicity is taken
into account with the potential $\pm\Delta$ on the two sublattices. With a
highly polarizable substrate, there is additional electron-phonon
interaction between the electrons in the film and phonons in the substrate, which may be long
range (i.e. momentum dependent). A Hamiltonian with these properties
has the form,
\begin{equation}
H = H_{\rm tb} + H_{\rm el-ph} + H_{\rm ph}
\end{equation}
where $H_{\rm tb}$ is the tight binding Hamiltonian representing the kinetic energy of the electrons hopping in the monolayer (note that there is no hopping perpendicular to the monolayer), $H_{\rm el-ph}$
describes the electron-phonon interaction, and $H_{\rm ph}$ is the
energy of the phonons in the substrate (treated as harmonic
oscillators, and including both kinetic and potential energy of the
ions).

The tight binding part of the Hamiltonian is written,
\begin{eqnarray}
H_{\rm tb} & = \sum_{\kvec\sigma}( & \phi_{\kvec} a^{\dagger}_{\kvec\sigma}c_{\kvec\sigma} + \phi^*_{\kvec} c^{\dagger}_{\kvec\sigma}a_{\kvec\sigma} \nonumber\\
&& + \Delta(a^{\dagger}_{\kvec\sigma}a_{\kvec\sigma} - c^{\dagger}_{\kvec\sigma}c_{\kvec\sigma})).
\end{eqnarray}
The first part represents the kinetic energy, where
$\phi_{\kvec}=-t\sum_i \exp(i\kvec.\deltavec_i)$, $t$ is the tight
binding parameter representing hopping between sites and $\deltavec_i$
are the nearest neighbor vectors from A to B sub-lattices,
$\deltavec_1=\tilde{a}(1,\sqrt{3})/2$,
$\deltavec_2=\tilde{a}(1,-\sqrt{3})/2$ and
$\deltavec_3=(-\tilde{a},0)$ and $\tilde{a}$ is the spacing between
carbon atoms in the plane (the tilde is used to avoid confusion with
the creation and annihilation operators). Electrons with momentum
$\kvec$ are created on A sites with the operator $a^{\dagger}_{\kvec}$
and B sites with $c^{\dagger}_{\kvec}$. The second part represents the
interaction between electrons in the monolayer and a static potential,
either induced by the substrate (in the case of graphene) or by
ionicity (in monolayers of III-V semiconductors). Here, A sites have a
higher potential, $\Delta$ and B sites are lower in energy by
$-\Delta$.  Breaking the symmetry between A and B sites in the
bipartite honeycomb lattice gives rise to a gap.

The phonon part of the Hamiltonian is,
\begin{equation}
H_{\rm ph} = \sum_{\qvec,z} \Omega_{\qvec,z} (b^{\dagger}_{\qvec,z} b_{\qvec,z} + d^{\dagger}_{\qvec,z} d_{\qvec,z})
\end{equation}
where phonons with momentum $\qvec$ are created in layer $z$ on A and
B sites with $b^{\dagger}_{\qvec,z}$ and $d^{\dagger}_{\qvec,z}$
respectively. Typically, the phonon dispersion, $\Omega_{\qvec}$ is
taken to be momentum independent as a good approximation to optical
phonons.  Typical phonon frequencies vary from 10s to 100s of meV.
For example, in BN phonon energies range from 110meV for transverse
acoustic phonons at the K point of the Brillouin zone to $200$meV for
optical phonons \cite{serrano2007a}. Due to ionicity, sites have a net
charge, so strong coupling between electrons and phonons is expected.

Finally, the interaction between electrons in the monolayer and phonons in the substrate (or superstrate in the case of graphene on a substrate) is,
\begin{eqnarray}
& H_{\rm el-ph} = \sum_{\kvec\qvec,z} \left[g^{(AA)}_{\qvec,z}a^{\dagger}_{\kvec-\qvec}a_{\kvec}(b^{\dagger}_{\qvec,z} + b_{-\qvec,z}) + g^{(BB)}_{\qvec,z} c^{\dagger}_{\kvec-\qvec}c_{\kvec}(d^{\dagger}_{\qvec,z} + d_{-\qvec,z})\right]\nonumber\\
& +\sum_{\kvec\qvec,z} \left[g^{(AB)}_{\qvec,z} a^{\dagger}_{\kvec-\qvec}a_{\kvec}(d^{\dagger}_{\qvec,z} + d_{-\qvec,z}) + g^{(BA)}_{\qvec,z}c^{\dagger}_{\kvec-\qvec}c_{\kvec}(b^{\dagger}_{\qvec,z} + b_{-\qvec,z})\right]
\label{eqn:ephint}
\end{eqnarray}
where the momentum-space coupling constants $g^{(XY)}_{\qvec,z}$
represent interactions between electrons in the film on sub-lattice $X$ and
phonons in the substrate on sub-lattice $Y$, and are defined as:
\begin{equation}
g^{(AA)}_{\kvec,z}=g^{(BB)}_{\kvec,z}=\sum_{i}e^{i\kvec\cdot\Rvec_i}g^{(z)}_{\zerovec}(\Rvec_{i})
\end{equation}
\begin{equation}
g^{(AB)}_{\kvec,z}=\sum_{i}e^{i\kvec\cdot(\Rvec_i+\ivec
  \tilde{a})}g^{(z)}_{\zerovec}(\Rvec_{i}+\ivec\tilde{a})
\end{equation}
and
\begin{equation}
g^{(BA)}_{\kvec,z}=\sum_{i}e^{i\kvec\cdot(\Rvec_i-\ivec
  \tilde{a})}g^{(z)}_{\zerovec}(\Rvec_{i}-\ivec\tilde{a}).
\end{equation}
Here, the
lattice vectors are $\Rvec_i=\tilde{n}\avec_{1}+\tilde{m}\avec_{2}$,
$\avec_{1}=3\ivec/2+\sqrt{3}\jvec/2$ and
$\avec_{2}=3\ivec/2-\sqrt{3}\jvec/2$.

The lattice Fr\"ohlich electron-phonon interaction used here has a
position space form,
\begin{equation}
g^{(z)}_{\mvec}(\nvec) = \kappa \exp(-|\nvec-\mvec|/R_{sc})[(\nvec-\mvec)^2+(\tilde{c}+z\tilde{a}_{\rm sub})^2]^{-3/2},
\label{eqn:frohlichinteraction}
\end{equation}
has been proposed for layered quasi-2D systems \cite{alexandrov2002a},
where $\kappa$ is a coupling constant. In
Eqn. \ref{eqn:frohlichinteraction}, $\nvec$ is the position of
electrons and $\mvec$ is the position of vibrating ions. Experiment
has demonstrated that this form explains interactions between
electrons in carbon nanotubes placed on SiO$_2$
\cite{steiner2009a}. The screening radius, $R_{sc}$ controls the
length scale of the interaction.  $\tilde{c}$ is the distance between
the graphitic thin film and surface atoms in the substrate.  In the
following, I take $\tilde{c}^2=2\tilde{a}^2$, since the distance
between graphene and substrate (which are typically bound by van der
Walls interactions) is likely to be slightly larger than between the
very strongly bound carbon atoms in the graphene layer. Ionic,
graphitic materials may bind more strongly to appropriate ionic
substrates leading to shorter $\tilde{c}$, which in this work is
represented (in combination with screening effects) with a reduced
$R_{sc}$. In practice, the effects of changing $\tilde{c}$ and
$R_{sc}$ on the form of the effective electron-electron coupling
mediated by phonons are very similar. Typically, this interaction is
with the surface ions only. The possibility of interactions with
addional layers in the bulk of the substrate can also be considered,
by adding a distance $z\tilde{a}_{\rm sub}$ to $\tilde{c}$, where $z$
is an integer, and then summing over all $z$ when calculating the
effective electron-phonon interaction (see
Sec. \ref{sec:selfconeph}). In this work, I set $\tilde{a}_{\rm
  sub}=\tilde{a}$ for convenience. The effect of adding interactions
with additional layers will be seen as a slight increase in the
effective interaction length. I will also consider the possibility of
having separate coupling constants for electrons on A and B
sites. Extensions to the formalism to allow this will be detailed in
Sec. \ref{sec:diffeph}.

The physical content of the electron-phonon interaction in
Eqn. \ref{eqn:ephint} can be seen in position space. Fourier transforms
of the electron-phonon interaction terms in the Hamiltonian
have the form, $H_{el-ph}\propto\sum_{\nvec,\mvec}
g_{\mvec}(\nvec)n_{\nvec} x_{\mvec}$ (since
$d^{\dagger}_{\mvec}+d_{\mvec}\propto x_{\mvec}$). Therefore, it can
be seen that the presence of electron density in the graphene sheet
leads to displacements in ion coordinates in the substrate. By
modifying the value of $R_{sc}$ it is possible to change the type of
interaction: For $R_{sc}\rightarrow 0$, the fully local Holstein
interaction, $H_{\rm Holstein}\propto g\sum_{i}n_{i}x_{i}$ is
recovered \cite{holstein1957a}. In the opposing limit,
$R_{sc}\rightarrow \infty$, the long range lattice Fr\"ohlich
interaction is recovered.

This section finishes with a note that the model used here has some
similarities to the ionic Hubbard model \cite{ionichubbard}. In the
ionic Hubbard model, the ionicity (introduced by an analogous
parameter $\Delta$) acts against the Mott insulating state (which is
caused by the repulsive Hubbard $U$). In contrast, in the model here,
the parameter $\Delta$ acts with the electron-phonon coupling to form
a charge density wave (CDW) insulating (gapped) state.

\section{Method}
\label{sec:method}

\begin{figure*}
\includegraphics[width=120mm]{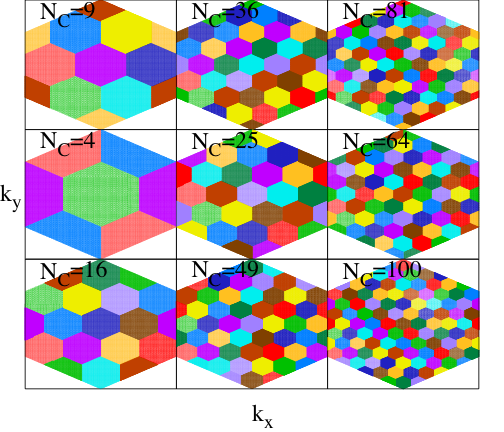}
\caption{(Colour online) Dynamical cluster approximation (DCA)
  sub-zones for cluster sizes of up to $N_C=100$. The axes show the
  $x$- and $y$-components of the momentum, $k_x$ and $k_y$. Within
  each sub-zone, the self-energy is taken to be momentum
  independent. This allows Green functions to be calculated in the
  thermodynamic limit, and convergence properties are particularly
  good as $N_C$ is increased. Note that the only symmetry taken into
  account is translation in $\kvec$-space.}
\label{fig:kpoints}
\end{figure*}

The electron-phonon Hamiltonian described above is extremely difficult
to solve exactly using numerical methods. An approximate solution can
be made using iterated perturbation theory within the dynamical
cluster approximation formalism.  The dynamical cluster approximation
(DCA) \cite{hettler1998a,hettler2000a} is one of the possible ways of
extending dynamical mean-field theory \footnote{Where DMFT is used to
  approximate low dimensional systems, it is often known as the local
  approximation} (DMFT)\cite{georges1996a} so that it can be applied
accurately to low dimensional systems. The Mermin-Wagner-Hohenberg
theorem indicates that the significant non-local fluctuations found in
some one- and two- dimensions could potentially lead to qualitatively
incorrect results from mean-field theories
\cite{mermin,hohenberg}. Moreover, DMFT has trouble dealing with the
spatial variations involved with interactions that extend over more
than one lattice site. DCA resolves this problem by developing a
mean-field theory around a cluster, rather than a single site,
therefore allowing the possibility of fluctuations or static spatial
variations up to the length scale of the cluster.

When applying DCA, the Brillouin zone is divided up into $N_C$
sub-zones centered about a momentum vector $\Kvec_{i}$ (see
Fig. \ref{fig:kpoints}) consistent with the symmetry of the whole
system. Within each sub-zone, the self-energy is approximated as a
momentum-independent function, so the Green function can be coarse
grained by integrating over the sub-zone,
\begin{eqnarray}
\Gbold(\Kvec_i , z) & = & \sum_{\kvec\in \Kvec_i} [\Ibold(z+\mu)+\Delta\sigmabold_3-\Phibold_{\kvec}-\Sigmabold(\Kvec_i , z)]^{-1}\label{eqn:greenfncoarse}\\
&\equiv& \left[\begin{array}{cc}G_{AA} & G_{AB} \\ G_{BA} & G_{BB}\end{array}\right]
\end{eqnarray}
where A and B represent sublattices and
\begin{equation}
\Phibold_{\kvec} = \left[\begin{array}{cc} 0 & \phi_{\kvec} \\ \phi^{*}_{\kvec} & 0\end{array}\right], \sigmabold_3 = \left[\begin{array}{cc} 1 & 0 \\ 0 & -1\end{array}\right].
\end{equation}
I will discuss the procedure for introducing long-range
electron-phonon interactions in Sec. \ref{sec:selfconeph}.

In finite size techniques, the number of particles is related to the
number of momentum points used in the calculation of the self
energy. In contrast, the DCA coarse-graining step involves an infinite
number of momentum points, so the thermodynamic limit is satisfied for
any cluster size. In the context of the perturbation theory for the
Migdal--Eliashberg theory used here, DCA has particularly good
convergence properties in cluster size $N_C$, so in principle smaller
clusters can be used leading to a significant improvement in
computational efficiency \cite{hague2003a,hague2005a}. I note that
when the DCA cluster size, $N_{C}=1$, calculations correspond to DMFT.

In several previous studies, $N_C=4$ and $N_C=16$ DCA clusters have
been used to understand Hubbard interactions on hexagonal/triangular
lattices (see e.g. \onlinecite{lee2008a}). I briefly discuss the
subzone schemes for hexagonal lattices with larger $N_C$. For the
lattices used here, the simplest way of defining the sub-zone vectors
is: $\tilde{\kvec}_{1}= \kvec_1 / \sqrt{N_C}$, $\tilde{\kvec}_{2}=
\kvec_2 / \sqrt{N_C}$ with the vectors $\Kvec_{i} = n\tilde{\kvec}_{1}
+ m\tilde{\kvec}_2$ with $n$ and $m$ integers. Here, the reciprocal
lattice vectors are $\kvec_1 =
(2\pi/3\tilde{a},2\pi/\sqrt{3}\tilde{a})$ and $\kvec_2 =
(2\pi/3\tilde{a},-2\pi/\sqrt{3}\tilde{a})$. There are likely to be
other valid lattices that can also be used, where the lattice and
sub-lattice are oriented at different angles. However, the lattices
used there are the simplest to implement.

Even for the simple cases considered here, the resultant lattices can
be ordered into groups. Clusters with $N_C=(3n)^{2}$ ($N_C=9$,
$N_C=36$, $N_C=81$ etc.) have sub-zones centered on the K and K$'$
points (here $n\ge 1$ is an integer). Those with $N_c=(3n-1)^2$
($N_C=4$, $N_C=25$ and $N_c=64$) make a second set where 3 sub-zones
share a corner at the K and K$'$ points, and the third set of
$N_c=(3n+1)^2$ ($N_C=16,64,100$ etc.)  where 3 sub-zones share a
corner at the K and K$'$ points and an edge with the full Brillouin
zone. Since the self energies would be identical in the 3 zones around
the K and K$'$ points in the latter 2 cases, they will poorly describe
the physics at the K and K$'$ points (which is especially
important for graphene). This is why I use only the $(3n)^2$ series.

To establish which $\kvec$ point belongs to a sub zone, it is
sufficient to find the closest $\Kvec_i$ point corresponding to
the center of the sub-zone (subject to shifts of reciprocal lattice
vectors). The edges of the shapes defined in this way are the hexagons
in the figure.

\subsection{Self-consistent equations for the Fr\"ohlich interaction}
\label{sec:selfconeph}

\begin{figure}
\includegraphics[width=65mm]{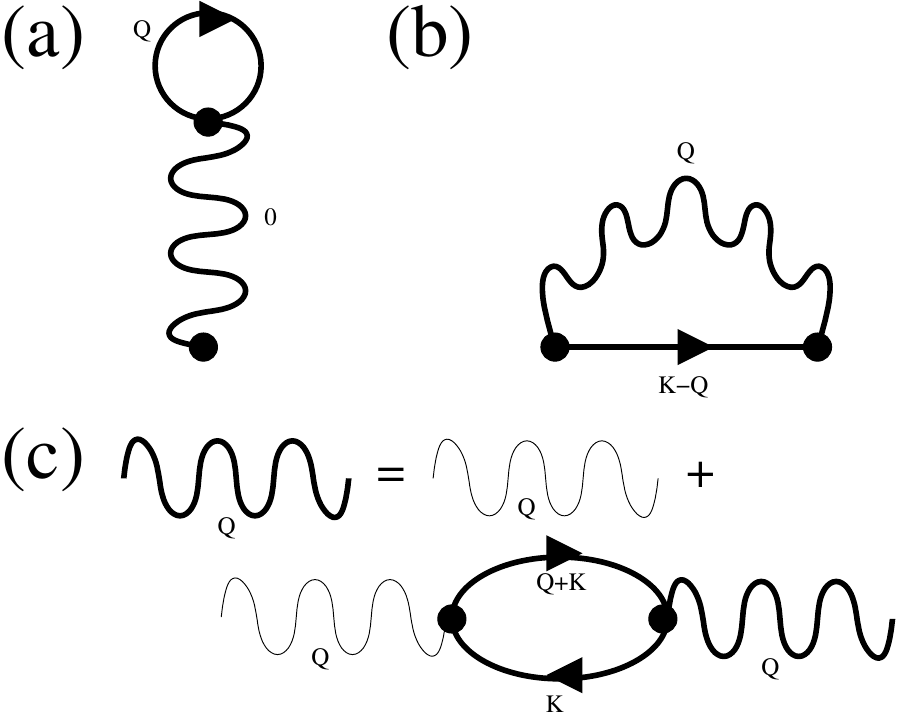}
\caption{Feynman diagrams showing the perturbation theory used in the
  work presented here. (a) Hartree diagram. For the symmetry broken
  states, this cannot be absorbed into the chemical potential, and is
  the main contributor to modification of the gap (b) The Fock
  potential is responsible for frequency dependence of the self-energy
  (c) Dyson equation for the phonon propagator. This renormalizes the
  phonon frequency, which can lead to further enhancement of the band
  gap.}
\label{fig:perturbation}
\end{figure}

I now describe how the perturbation theory for the long-range electron-phonon
interaction is used in conjunction with the DCA. The perturbation
theory used here can be seen in Fig. \ref{fig:perturbation}. Panel (a)
shows the Hartree diagram. For the symmetry broken states, this cannot
be absorbed into the chemical potential, and is the main contributor
to modification of the gap.  The Fock diagram (shown in panel (b))
is responsible for frequency dependence of the self-energy. Phonon
propagators are modified using a Dyson equation (panel (c)) which
modifies the phonon frequency, and can lead to further enhancement of
the band gap. Following the standard formulation of the DCA, where
momentum is not conserved at vertices within the sub-zones (see
e.g. Ref. \onlinecite{maier2005a} for a good review), momentum sums in
the perturbation theory are reduced to sums over the average momenta
of the sub-zones. Therefore,
\begin{equation}
\Sigma^{(H)}_{XY}=-2\frac{T}{N_C}\delta_{XY}\sum_{lm\Kvec'}D_{Xl}(\zerovec)G_{ll}(\Kvec',\omega_m)
\label{eqn:hartse}
\end{equation}
\begin{equation}
\Sigma^{(F)}_{XY}(\omega_n\Kvec) = \frac{T}{N_C}\sum_{\omega_s,\Qvec} G_{XY}(\omega_n-\omega_s,\Kvec-\Qvec)D_{XY}(\omega_s\Qvec)
\label{eqn:fockse}
\end{equation}
where $\Qvec$ represent the centers of the coarse-grained cells for
phonon momenta.
\begin{equation}
\left[D^{-1}(\Qvec)\right]_{XY} = \left[d^{-1}(\Qvec)\right]_{XY}-\left[\Pi(\Qvec)\right]_{XY}
\label{eqn:phonprop}
\end{equation}
where
\begin{equation}
\Pi_{XY}(\Qvec,\omega_s) = -2\frac{T}{N_C}\sum_{\Kvec\omega_n} G_{XY}(\Qvec+\Kvec,\omega_n+\omega_s) G_{YX}(\Kvec,\omega_n)
\label{eqn:phonse}
\end{equation}
and the non-interacting phonon propagator is,
\begin{equation}
d_{XY}(\Qvec,\omega_s) = \lambda_{XY}(\Qvec)\Omega^2/(\Omega^2+\omega_s^2).
\label{eqn:phononprop}
\end{equation}
where $\lambda_{XY}(\Qvec)$ is the dimensionless electron phonon
coupling averaged to a single sub-zone centered around momentum
$\Qvec$.

It remains to define how to deal with the momentum dependent
electron-phonon coupling within the DCA formalism. Here the
dimensionless, momentum dependent electron-phonon coupling, $\lambda$,
is incorporated using the following procedure. I first note that in
position space, the standard dimensionless electron-phonon coupling is
defined to be, $\lambda=\sum_{\mvec
  z}|g^{(z)}_{\mvec}(0)|^2/t\hbar\Omega$ (this value is the ratio of
the polaron energy in the atomic limit to the hopping, $t$, see
e.g. Ref. \onlinecite{hague2007b} for more details), and that the
Fourier transform of this definition to convert the sum to momentum
space gives $\lambda=\sum_{\kvec, XY
  z}|g^{(XY)}_{\kvec,z}|^2/t\hbar\Omega$. Following this, I define the
coupling for a single $\kvec$ and $z$ value to be,
%
%
%
\begin{equation}
\tilde{\lambda}^{(XY)}_{\kvec,z}=|g^{(XY)}_{\kvec,z}|^2/t\hbar\Omega
\end{equation}

\begin{figure}
\includegraphics[width=65mm]{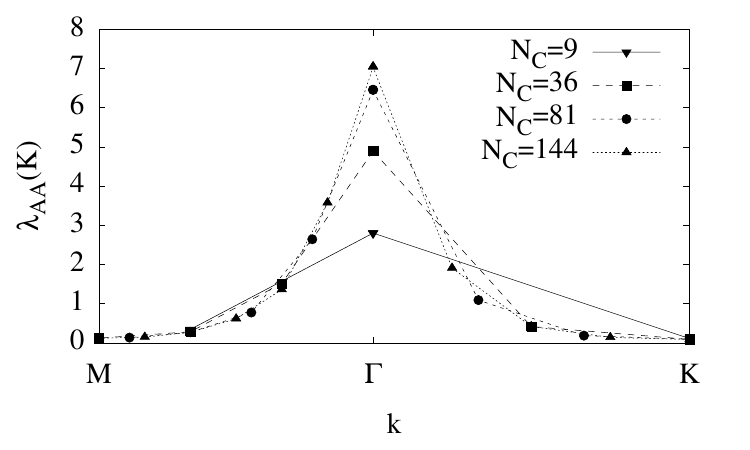}
\includegraphics[width=65mm]{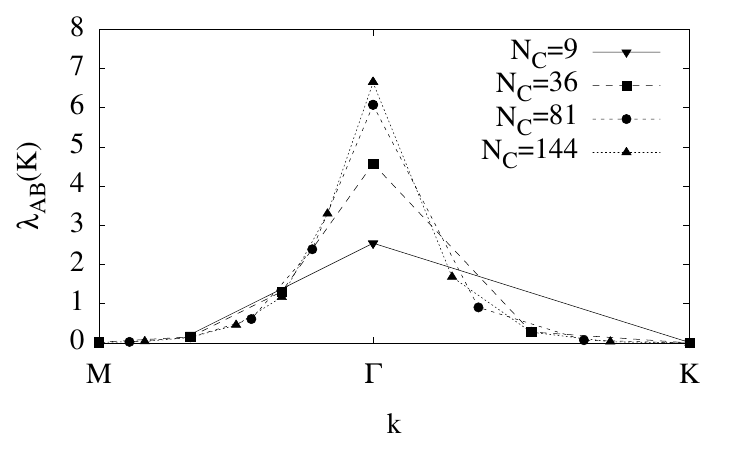}
\caption{Values of $\lambda_{XY}(\Kvec_i)$ plotted for zones that are
  centered on the high symmetry directions. The Fr\"ohlich interaction
  used has an average $\lambda$ value of 1, so due to the
  inhomogeneity of the interaction across the Brillouin zone, the
  interaction in unit cells close to the $\Gamma$ point is larger than
  1, and smaller than 1 elsewhere. The peak is supressed for small
  clusters due to averaging across DCA sub-zones, but reaches the peak
  value for cluster sizes of order 144. The difference between
  $\lambda_{AA}(\Kvec)$ and $\lambda_{AB}(\Kvec)$ at $\Gamma$ is
  important for the effective interaction strength in the Hartree
  diagram.}
\label{fig:lambdazero}
\end{figure}

The dimensionless electron-phonon coupling can then be related to the
value of $\lambda_{XY}(\Qvec)$ used in equation \ref{eqn:phononprop}
via,
\begin{equation}
\lambda_{XY}(\Kvec_i)=2N_{C}\lambda \frac{\sum_{\kvec\in\Kvec_{i},z=1}^{z=N_{Z}}\tilde{\lambda}^{(XY)}_{\kvec,z}}{\sum_{\kvec'\in BZ,\alpha\beta,z=1}^{z=N_{Z}}\tilde{\lambda}^{(\alpha\beta)}_{\kvec',z}}
\label{eqn:lambdacoarsegrain}
\end{equation}
where $N_{Z}$ is the number of planes of vibrating ions in the bulk substrate
that electrons in the plane are coupled to (note that the electrons do
not hop into the substrate). The reason for defining $\lambda_{XY}$ in this
way is that it is a convenient way of cancelling the non-standard
coupling constant, $\kappa$, and replacing it with the standard
dimensionless electron-phonon coupling, $\lambda$. In this expression,
the sum in the denominator leads to an average value that is
proportional to $\lambda$ multiplied by the number of lattice sites,
$2N_{C}$ (there are 2 sub-lattices for every cluster site), so by
multiplying the average value of $\tilde{\lambda}_{\kvec,z}$ in each
DCA sub-zone by $2N_{C}\lambda/\sum_{\kvec'\in
  BZ,nm,z}\tilde{\lambda}^{(nm)}_{\kvec',z}$, factors of $\kappa$
cancel. To give an idea about how lambda varies for different DCA
subzones, $\lambda_{XY}(\Kvec_i)$ is plotted in
Fig. \ref{fig:lambdazero} for zones that are centered on the high
symmetry directions.

Self-consistency is then carried out as follows:
\begin{enumerate}
\item Initialise by calculating the coarse-grained electron-phonon
  interaction from Eqn. \ref{eqn:lambdacoarsegrain}, and the Green
  function from Eqn. \ref{eqn:greenfncoarse} with the electron self-energy set to zero.
\item Calculate the phonon self-energy (polarization bubble) from Eqn. \ref{eqn:phonse} 
\item Calculate the renormalised phonon propagator from Eqn. \ref{eqn:phonprop}.
\item Use the renormalized phonon propagator to calculate the Hartree and Fock contributions to the electron self-energy following Eqns. \ref{eqn:hartse} and \ref{eqn:fockse}.
\item Re-calculate the Green function from Eqn. \ref{eqn:greenfncoarse}.
\item Repeat steps 2-5 until converged.
\end{enumerate}

The resulting formalism is quite robust for phonon energies that are
of the order of, or smaller than $kT$, as is the case with all
calculations made here for room temperature and phonons with energies
in the range $10-100$meV, since the phonon propagator acts like a
$\delta$-function when $\hbar\Omega<k_{B}T$. Since the Hartree diagram
does not have any frequency sums that include the phonon propagator,
the sum over Matsubara frequencies in the next most important diagram
(the Fock diagram) is severely truncated, leading to a much reduced
contribution (with the caveat that the Green function must have small
values for low Matsubara frequencies, which is ensured by the V-shaped
form of the density of states in graphene). In practice, this means
that all other terms in the perturbation expansion for the electron
self-energy will be very much smaller, and that in this case (because of
the vanishing DOS at the Fermi surface) Migdal's theory holds. Similar
considerations apply for the phonon self-energy such that the single
polarisation bubble formed from dressed electron propagators should be
the dominant term in the perturbation expansion. Therefore, the
approximation used here is expected to be highly accurate for large
cluster sizes.


\subsection{Extensions for different interactions on A and B sublattices}
\label{sec:diffeph}

Finally, I note that it is possible to have different electron-phonon
interactions on each of the A and B sub-lattices. This may occur since
atoms on the A and B sites are different, so that the orbitals holding
the electrons that cause the ion displacements have a different
form. In practice, I would expect this effect to be quite small
($\lesssim 30\%$) if A and B sites are in the same period of the periodic
table, but this effect may be larger if the atoms come from different
periods.

Starting again from the expression,
\begin{equation}
\tilde{\lambda}^{(XY)}_{\kvec,z}=|g^{(XY)}_{\kvec,z}|^2/t\hbar\Omega
\end{equation}
Two dimensionless constants can now be introduced,
$\lambda_{A}\propto\kappa_A^2$ and $\lambda_{B}\propto\kappa_B^2$. I
note that the values of $|g^{(XY)}_{\kvec,z}|^2$ are proportional to
$\lambda_A$ if both sublattices are of type A, $\lambda_B$ if both
sublattices are of type B, and $\sqrt{\lambda_{A}\lambda_B}$ if the
sublattices are different. It is worth noting at this stage that the
factor $\sqrt{\lambda_{A}\lambda_B}$ for off diagonal terms means that
the inter-site interactions are reduced faster than the simple average
of $\lambda_{A}$ and $\lambda_{B}$, which makes the interaction much
more localized if the difference between $\lambda_{A}$ or
$\lambda_{B}$ is significant.

The dimensionless electron-phonon coupling can then be related to the
value of $\lambda_{XY}(\Qvec)$ used in the self-consistent equations
via,
\begin{equation}
\lambda_{XY}(\Kvec_i)=\frac{N_{C}}{2}(\sqrt{\lambda_{A}}+\sqrt{\lambda_{B}})^2 \frac{\sum_{\kvec\in\Kvec_{i},z}\tilde{\lambda}^{(XY)}_{\kvec,z}}{\sum_{\kvec'\in BZ,\alpha\beta,z}\tilde{\lambda}^{(\alpha\beta)}_{\kvec',z}}
\label{eqn:lambdacoarsegrainbb}
\end{equation}
The prefactor in this expression is different to the previous one, since the sum in the
denominator is proportional to
$\lambda_A+\lambda_B+2\sqrt{\lambda_{A}\lambda_B}=(\sqrt{\lambda_{A}}+\sqrt{\lambda_{B}})^2$.

\begin{figure*}
\includegraphics[width=65mm]{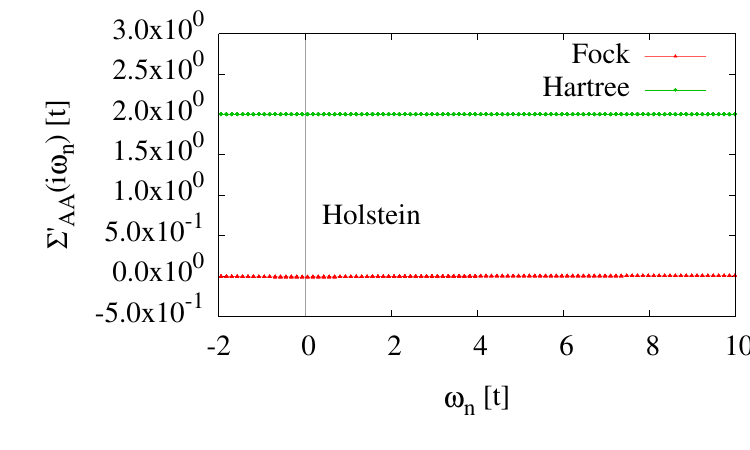}\includegraphics[width=65mm]{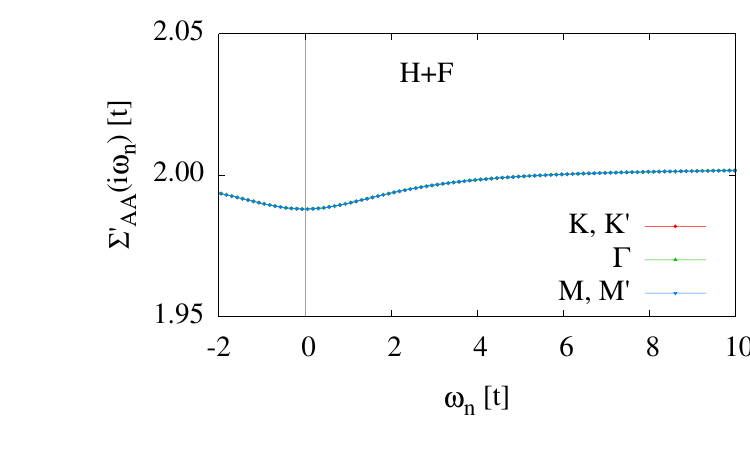}\\
\includegraphics[width=65mm]{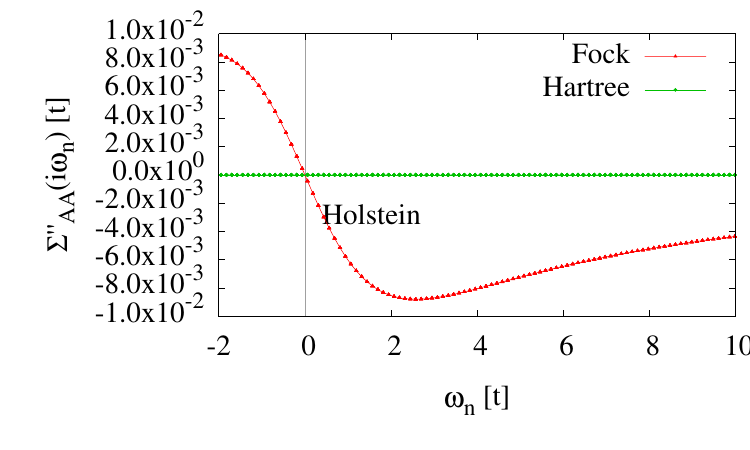}\includegraphics[width=65mm]{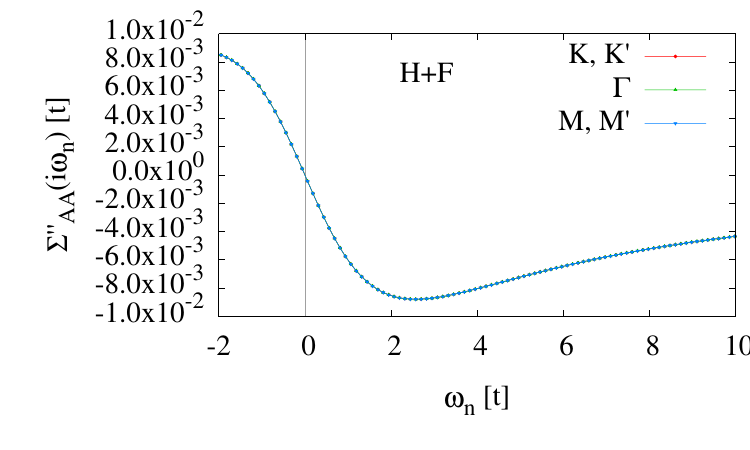}\\
\includegraphics[width=65mm]{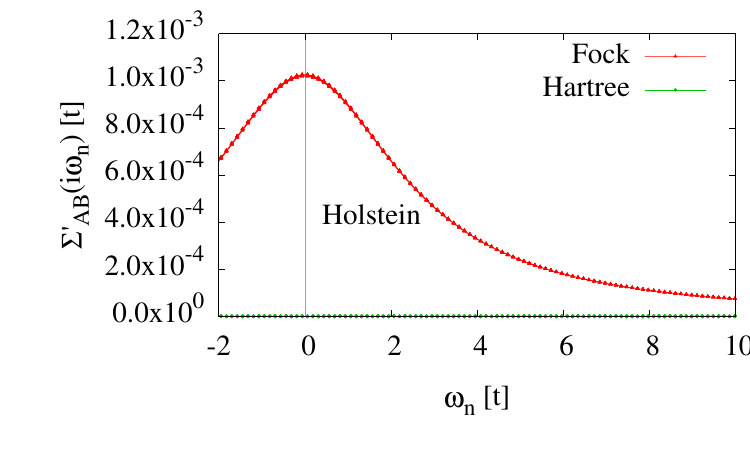}\includegraphics[width=65mm]{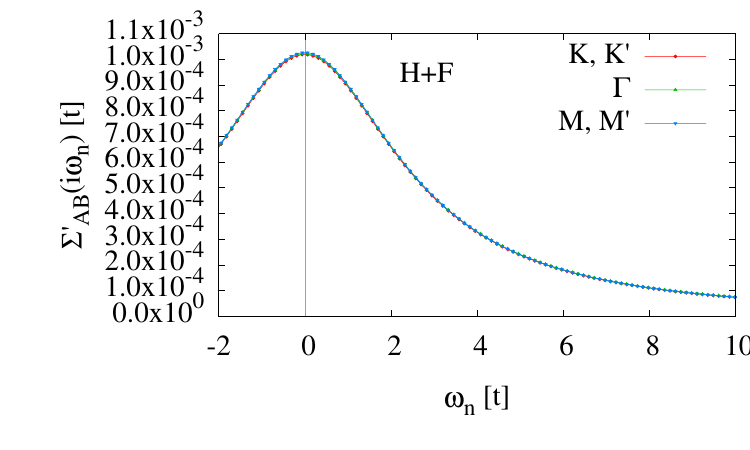}\\
\includegraphics[width=65mm]{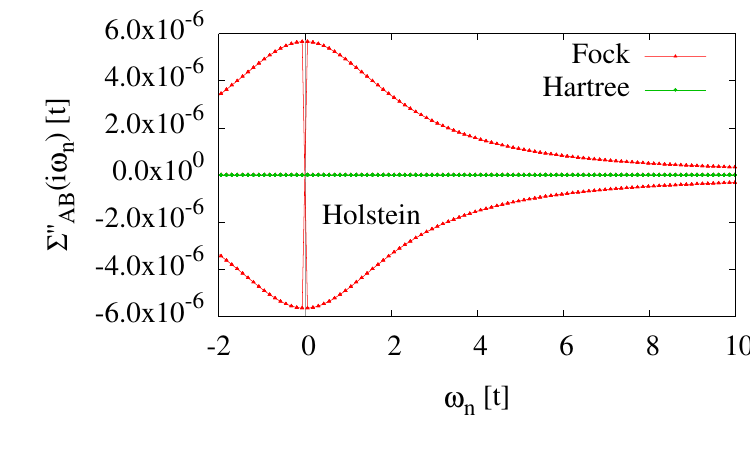}\includegraphics[width=65mm]{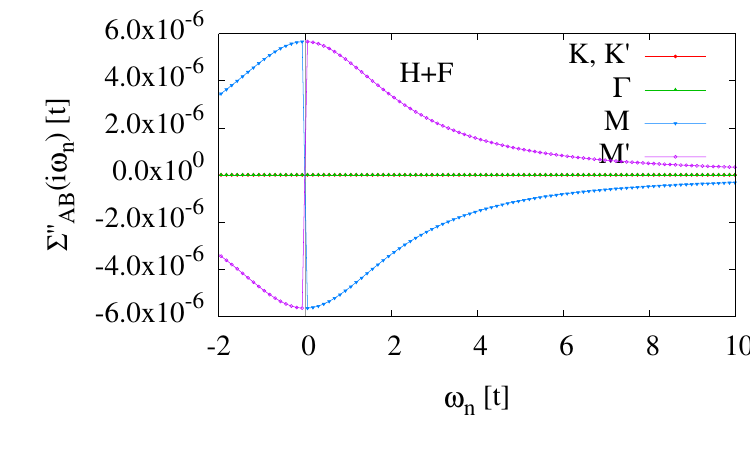}\\

\caption{(color online) Comparison of Hartree and Fock parts of self
  energy for the case of the Holstein interaction, $\Sigma = \Sigma' +
  i\Sigma''$ vs Matsubara frequency in a 9 site cluster. The largest
  element in the self energy matrix is the real part of the on-diagonal
  Hartree term, which is momentum independent, and it is this value
  that defines the size of the gap. Panels on the left show individual
  Hartree and Fock contributions to the gap and on the right show the
  momentum dependence of the total self energy. $\Sigma_{AA}$ is
  essentially momentum independent (so in the upper panel, curves for
  K and $\Gamma$ points lie directly under those for the M point), and
  $\Sigma_{AB}'$ has weak momentum dependence, so the points can only
  be differentiated under high magnification. $\Sigma_{AB}''$ is very
  small. Primed points are related to unprimed ones by inversion
  around the origin. N.B. The Fock term is still the most important
  contribution in some cases, for example the inverse mass depends on
  derivatives of the self energy, so this will be given by the Fock
  term. The points marked M and M$'$ represent zones that border on
  the M point (but where the DCA sub-zone center in the $N_{C}=9$
  cluster is offset slightly from the M point so that the values can
  be distinct). $T=0.02t$, $\Delta = 0.1t$, $\Omega=0.01t$ and
  $\lambda=2$.}
\label{fig:cmpse}
\end{figure*}

\begin{figure*}
\includegraphics[width=65mm]{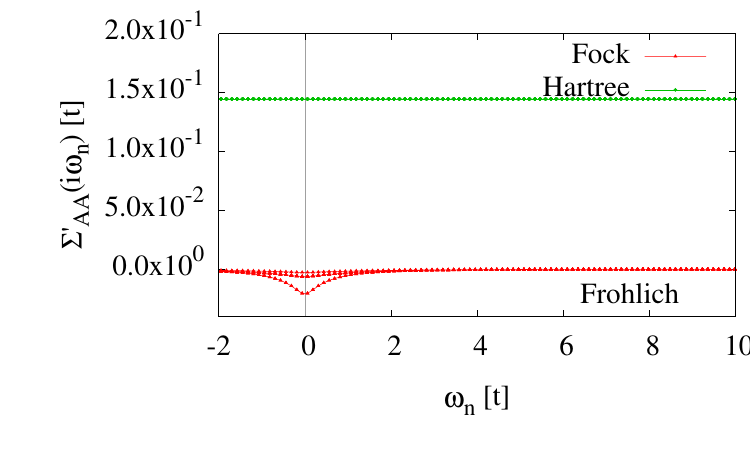}\includegraphics[width=65mm]{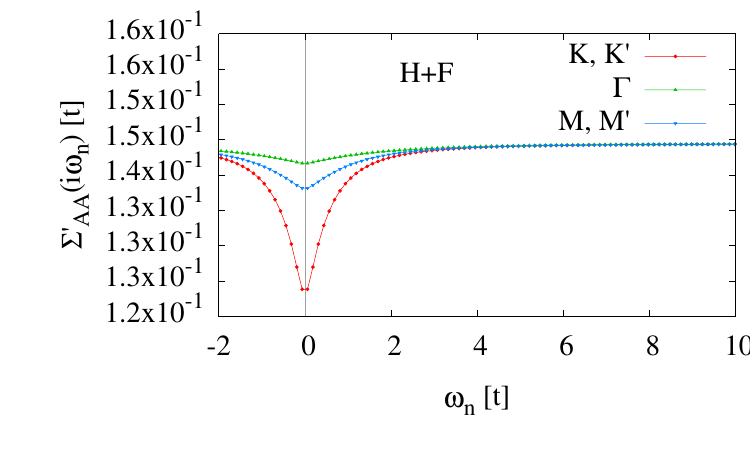}\\
\includegraphics[width=65mm]{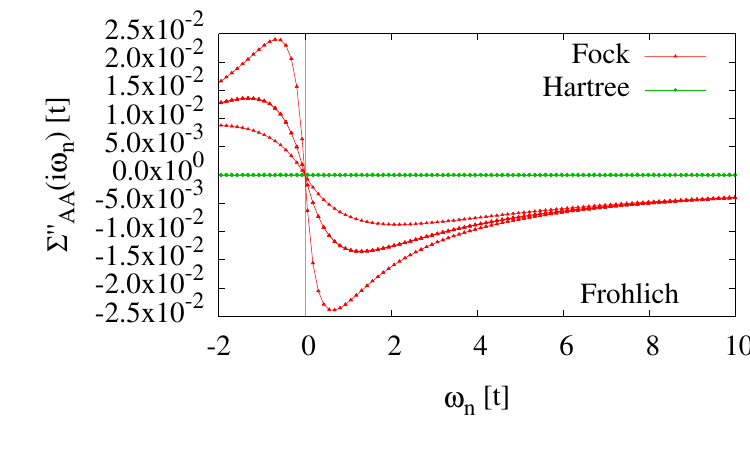}\includegraphics[width=65mm]{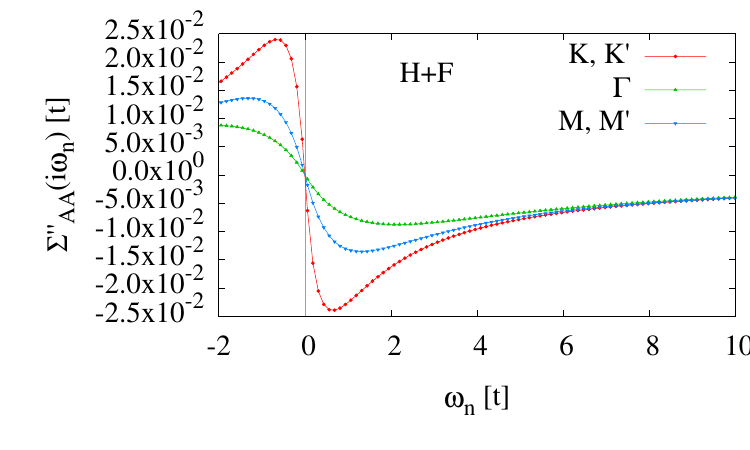}\\
\includegraphics[width=65mm]{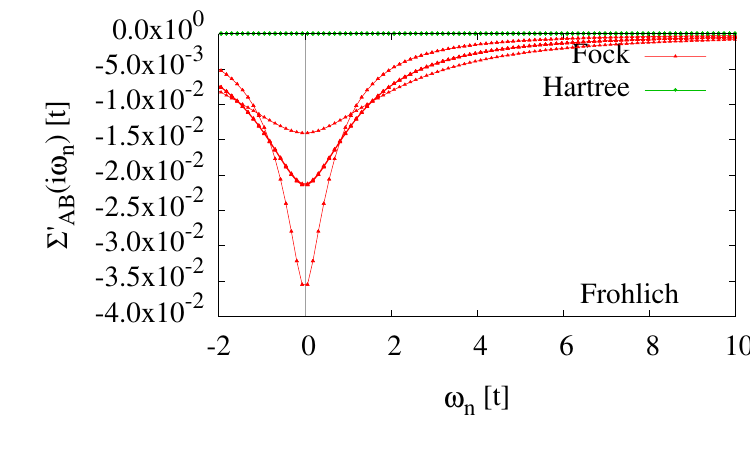}\includegraphics[width=65mm]{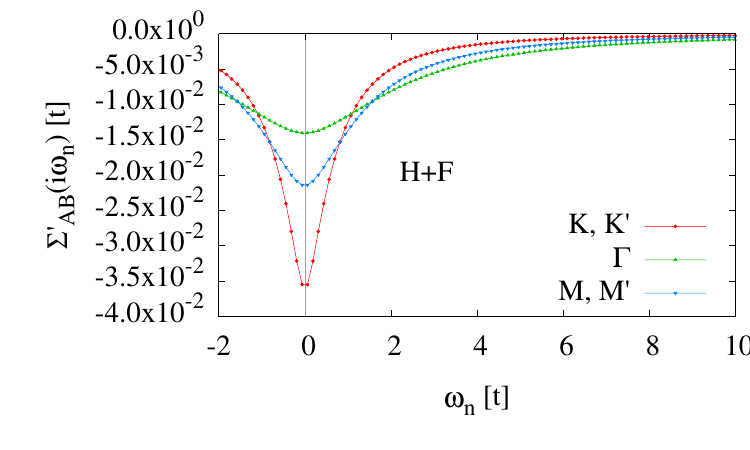}\\
\includegraphics[width=65mm]{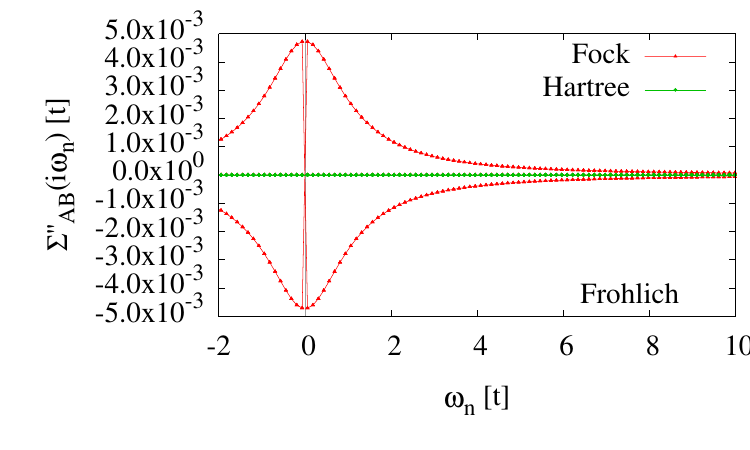}\includegraphics[width=65mm]{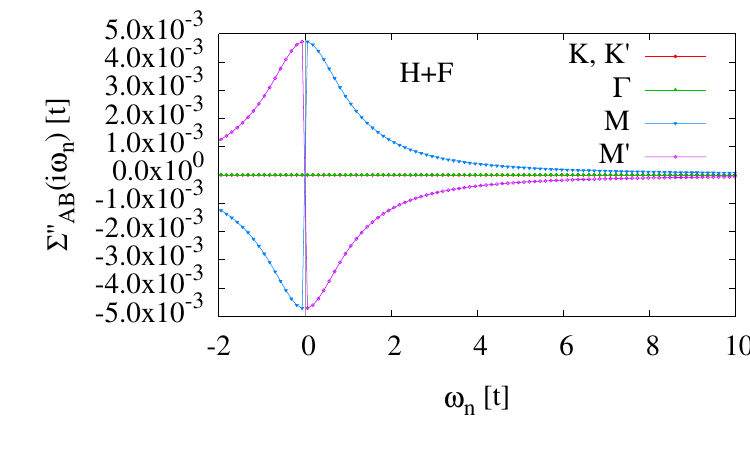}\\

\caption{(color online) Comparison of Hartree and Fock parts of self
  energy for a Fr\"ohlich interaction, $\Sigma = \Sigma' + i\Sigma''$
  vs Matsubara frequency. The on-diagonal contributions can be seen at
  the top and the off-diagonal ones at the bottom. Again, the largest
  contribution to the self energy matrix is the real part of the
  on-diagonal Hartree term, which defines the gap and is momentum
  independent. As in the Holstein case, the Hartree contibution to the
  off-diagonal self-energy is necessarily zero, and the Fock diagram
  is the largest off-diagonal contibution although it is still around
  1/4 of the magnitude of the on-diagonal Hartree term. The Fock term
  has a significantly bigger momentum dependence than in the Holstein
  case. N.B. The contributions to $\Sigma_{AB}''$ for K, K$'$ and
  $\Gamma$ points are all zero, so can not be distinguished from each
  other. Here, $T=0.02t$, $\Delta = 0.1t$, $\Omega=0.01t$, $\lambda=2$ and
  $N_C=9$. The Fock contributions become relatively smaller compared
  to the Hartree term as cluster size increases.}
\label{fig:cmpsefroh}
\end{figure*}

\section{Results}
\label{sec:results}


\begin{figure}
\includegraphics[width=65mm]{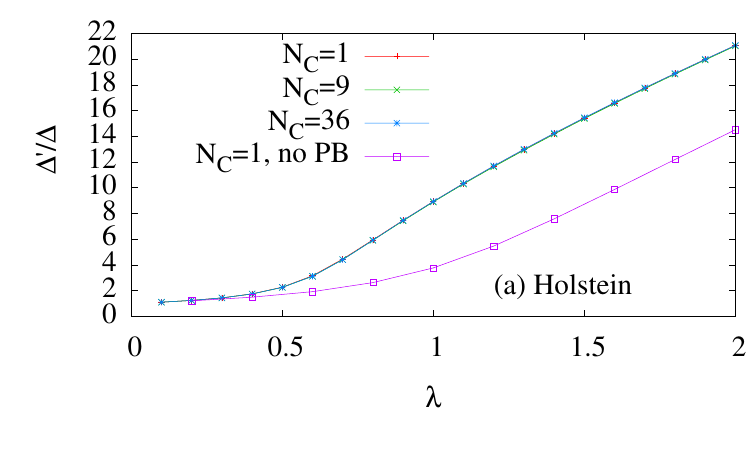}\\
\includegraphics[width=65mm]{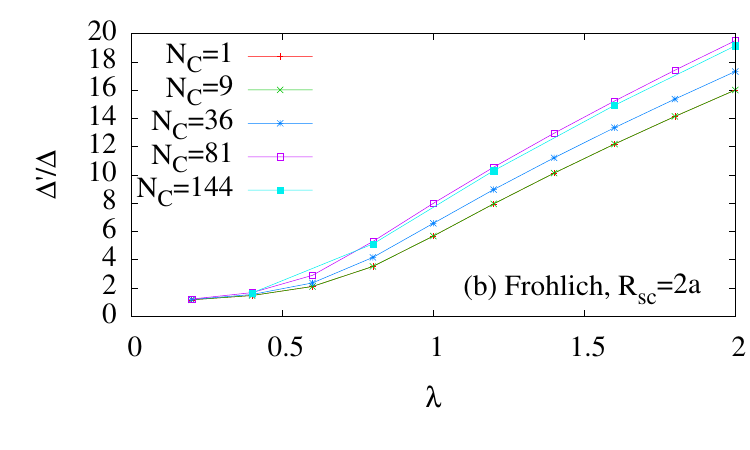}\\
\includegraphics[width=65mm]{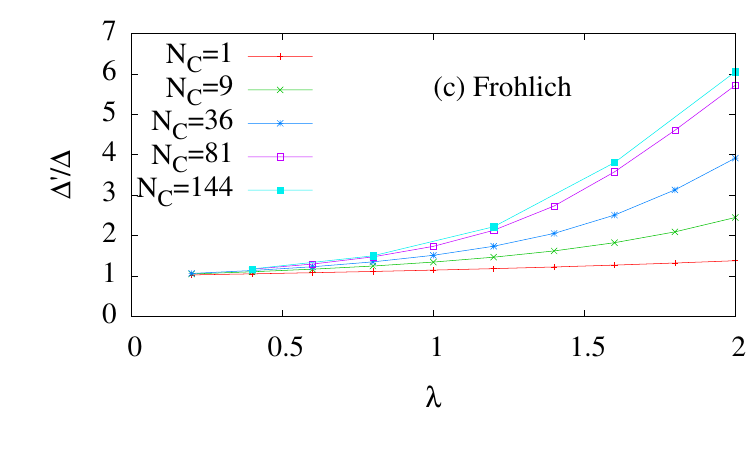}
\caption{(color online) (a) Gap enhancement $\Delta'/\Delta$ vs
  electron-phonon coupling $\lambda$ for a Holstein interaction with
  $\Delta/t=0.1$, comparing results from the dynamical mean-field
  theory (corresponding to $N_{C}=1$) and DCA. There are only small
  corrections due to momentum dependence. (b) Gap enhancement for the
  long range Fr\"ohlich interaction, $R_{sc}=2a$ (c) Gap enhancement
  for the long range Fr\"ohlich interaction. Here $T=0.02t$ and
  $\Omega=0.01t$. The initial increase in the gap as $N_{C}$ increases
  arises because the long range interaction is not homogenous across
  the Brillouin zone, so the effective value of $\lambda(\Qvec=0)$
  (that is relevant to the Hartree diagram) is larger than the average
  $\lambda$ in all cluster sizes except $N_C=1$. The enhancement is
  essentially converged for cluster sizes of $N_c=144$. (N.B. Since
  $\Delta=0.1t$, the gap enhancement is 10 times larger than the gap,
  $\Delta'$, which can be read from the real part of the self-energy
  at large Matsubara frequency.)}
\label{fig:gapenhancecombined}
\end{figure}

\begin{figure*}
\includegraphics[width=65mm]{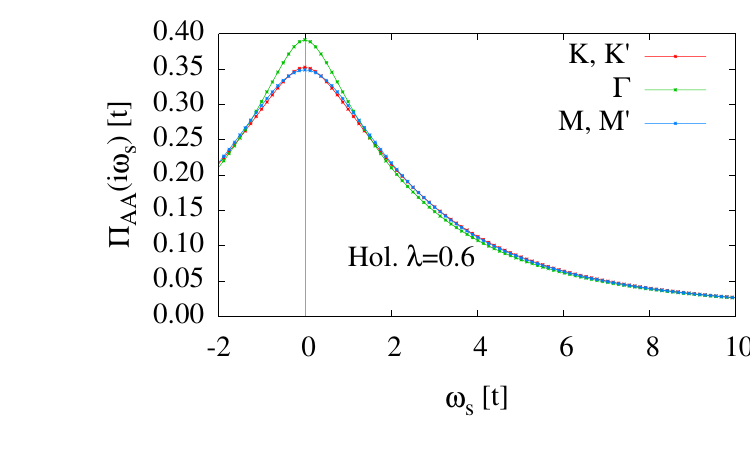}\includegraphics[width=65mm]{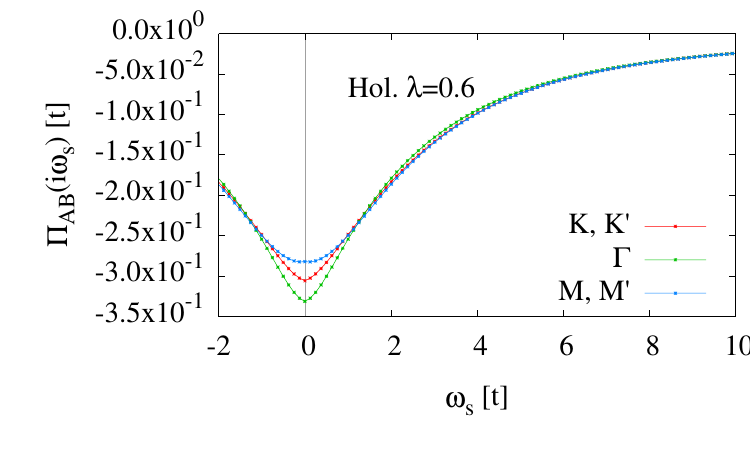}\\
\includegraphics[width=65mm]{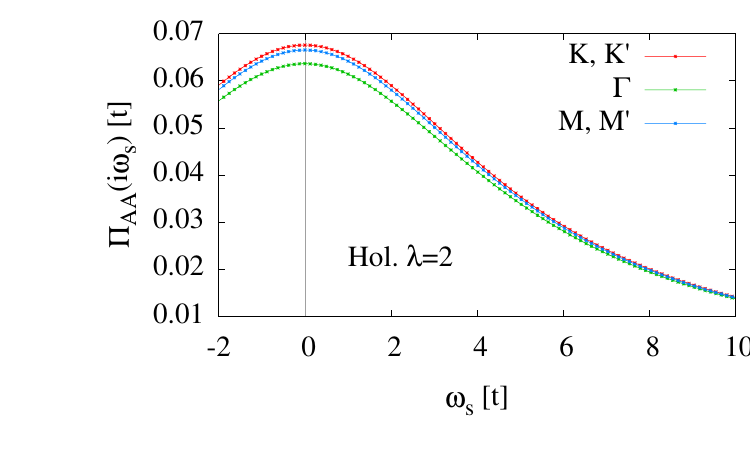}\includegraphics[width=65mm]{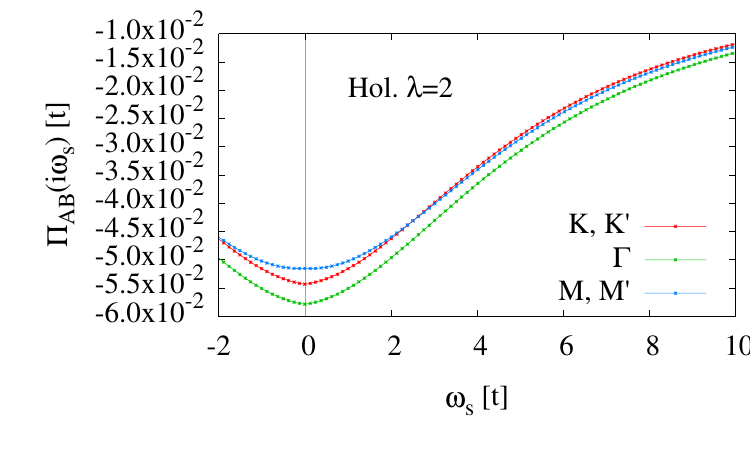}\\
\includegraphics[width=65mm]{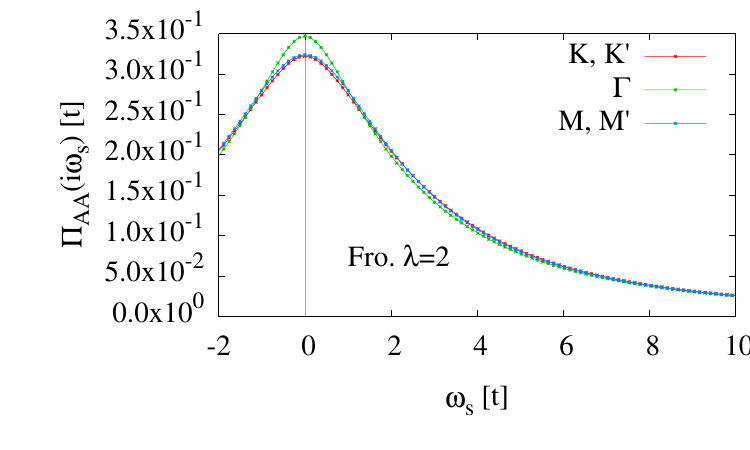}\includegraphics[width=65mm]{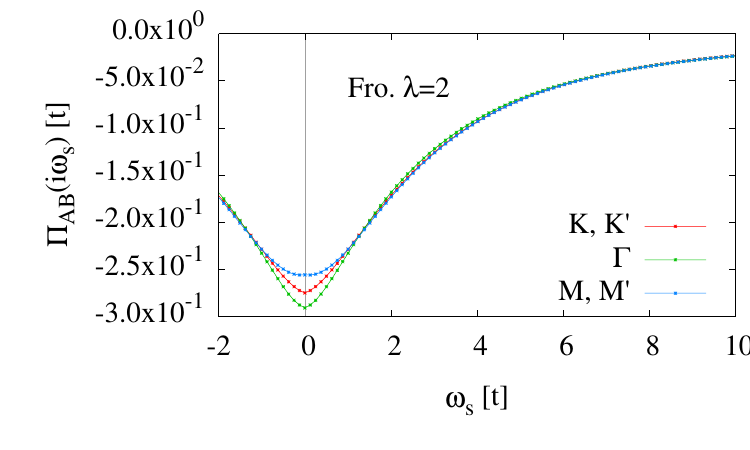}\\

\caption{(color online) Comparisons of the phonon self energy for
  Holstein and Fr\"ohlich interactions with various $\lambda$. As the
  interaction strength increases, the phonon self energy decreases due
  to the gap at the Fermi energy that reduces the value of the Green
  function at low Matsubara frequencies. The self energy is only
  weakly momentum dependent, but it is possible to discern the
  variation across the Brillouin zone.}
\label{fig:phonse}
\end{figure*}

\begin{figure}
\includegraphics[width=65mm]{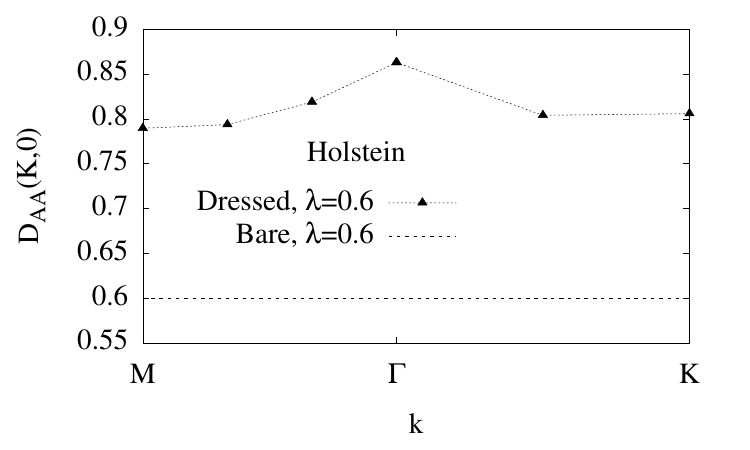}
\includegraphics[width=65mm]{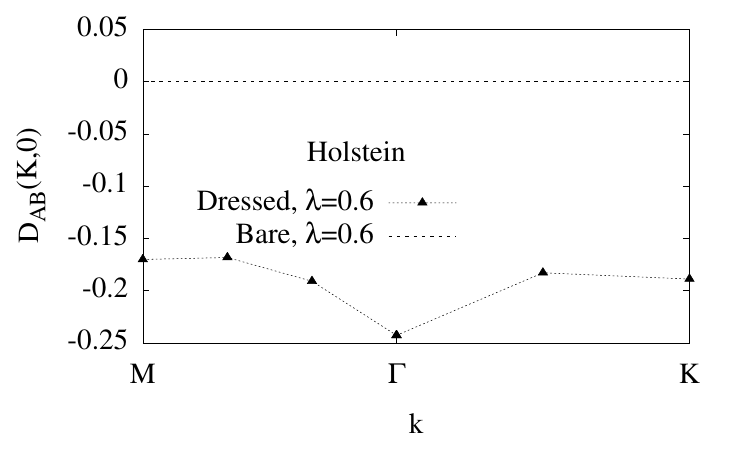}
\includegraphics[width=65mm]{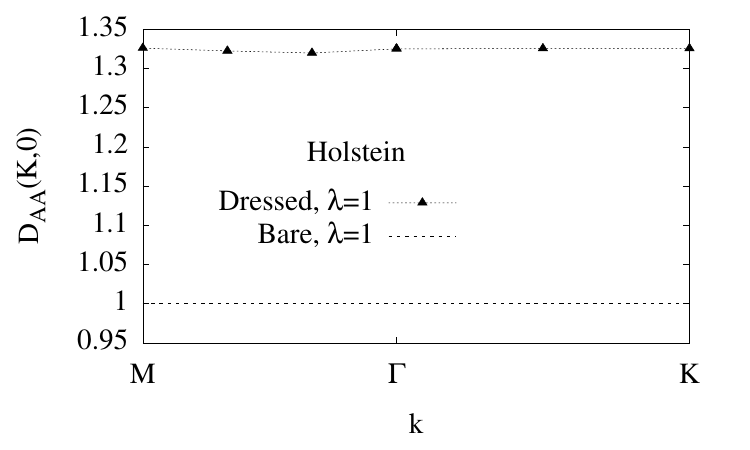}
\includegraphics[width=65mm]{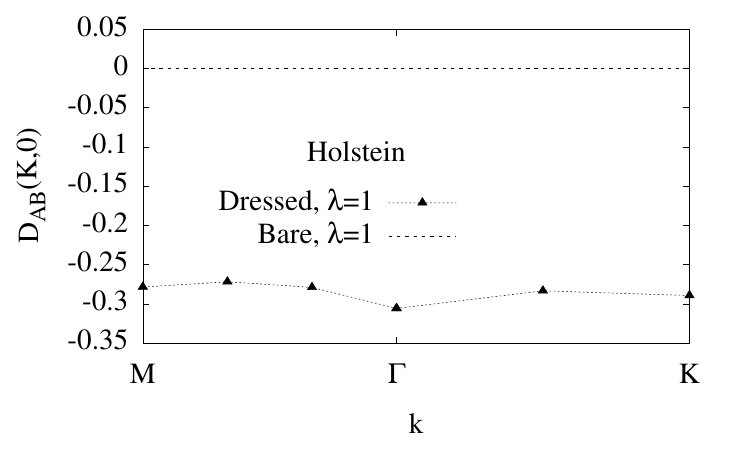}
\caption{Effective coupling in the Holstein model for a range of
  couplings, $D_{XY}(Q,0)$. The effective interaction becomes larger
  more rapidly than $\lambda$, and the form changes from momentum
  independent to weakly momentum dependent.}
\label{fig:effcouple}
\end{figure}

\begin{figure}
\includegraphics[width=65mm]{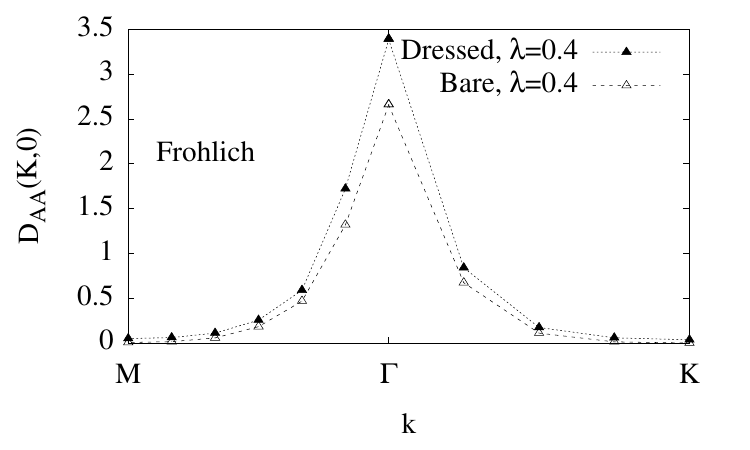}
\includegraphics[width=65mm]{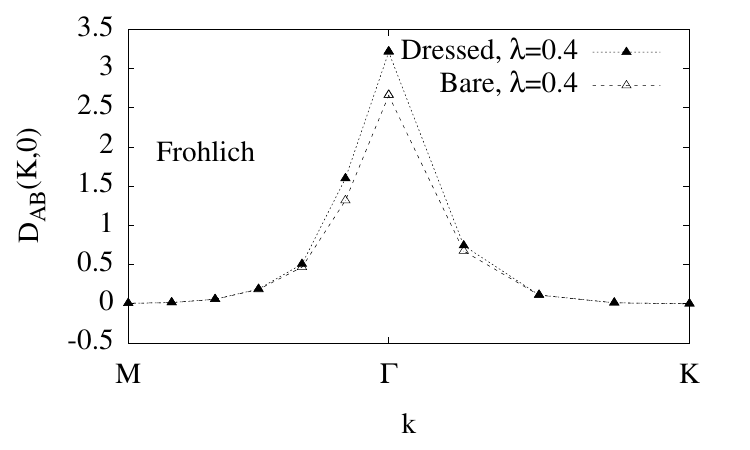}
\includegraphics[width=65mm]{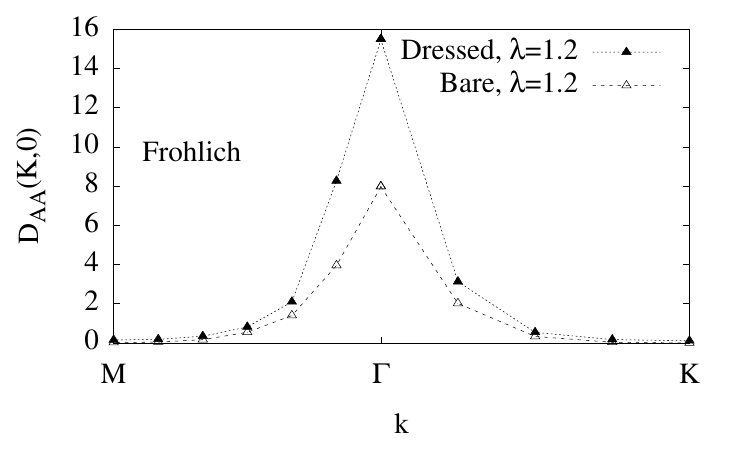}
\includegraphics[width=65mm]{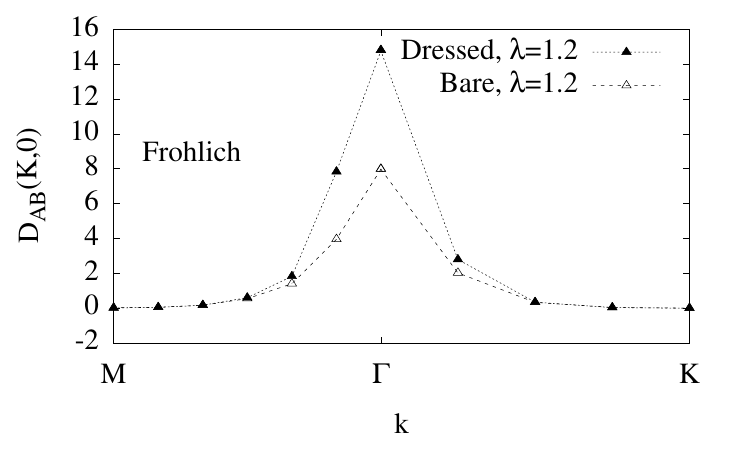}

\caption{Effective coupling in the Fr\"ohlich model for a range of
  couplings, $D_{ij}(Q,0)$. This property is strongly momentum
  dependent, and the momentum dependence becomes slightly larger as
  coupling increases. It is the effective coupling that leads to
  non-local position space variations that require large clusters to
  treat.}
\label{fig:effcouplefro}
\end{figure}

\begin{figure}
\includegraphics[width=65mm]{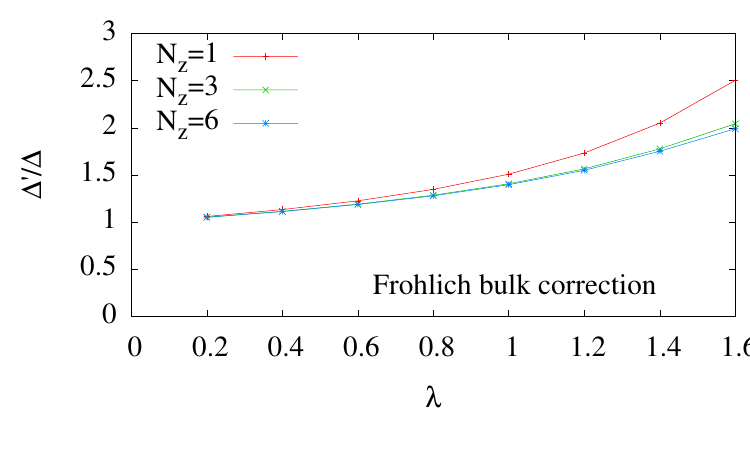}
\caption{(color online) Corrections taking account of the finite depth
  of the bulk substrate. The traces show the effect of interactions
  between electrons in the monolayer and vibrations in the surface
  atoms only ($N_{z}=1$) and interactions with vibrations in 3 and 6
  layers of bulk substrate respectively (N.B. electrons only hop in
  the monolayer, and there is no hopping into the bulk of the
  substrate). The effect of the bulk of the substrate is to reduce the
  enhanced gap by around 20\%.}
\label{fig:bulkcorr}
\end{figure}

\begin{figure}
\includegraphics[width=65mm]{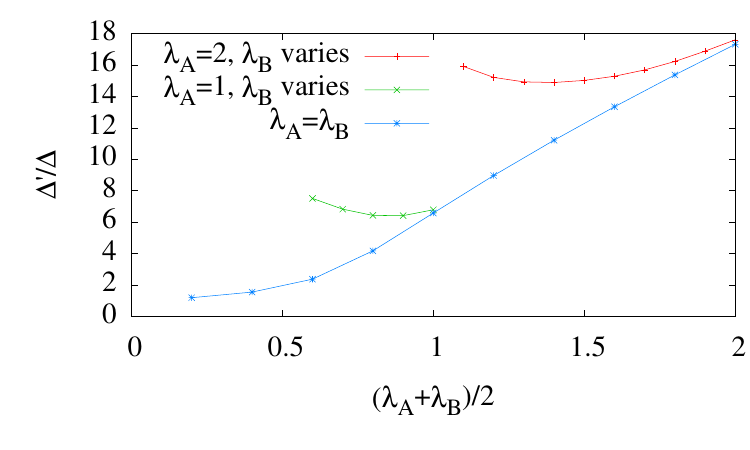}
\caption{(color online) The effect of taking
  $\lambda_{A}\neq\lambda_{B}$ when $R_{sc}=2a$ and $N_{C}=36$. Since electrons on
  different sub-lattices are contained in orbitals of different atoms,
  the interaction between electrons and phonons may depend on the
  sub-lattice. The enhancement is mainly determined by the largest of
  $\lambda_{A}$ or $\lambda_B$ with small changes to the overall
  enhancement as the other coupling is varied.}
\label{fig:lambdaneq}
\end{figure}

\begin{figure}
\includegraphics[width=65mm]{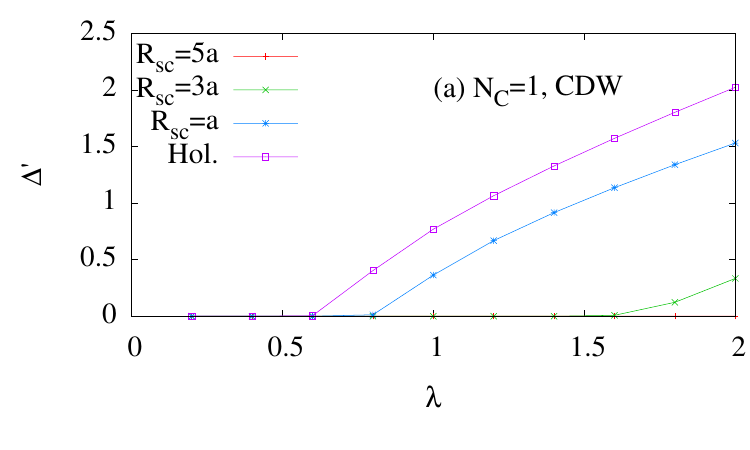}
\includegraphics[width=65mm]{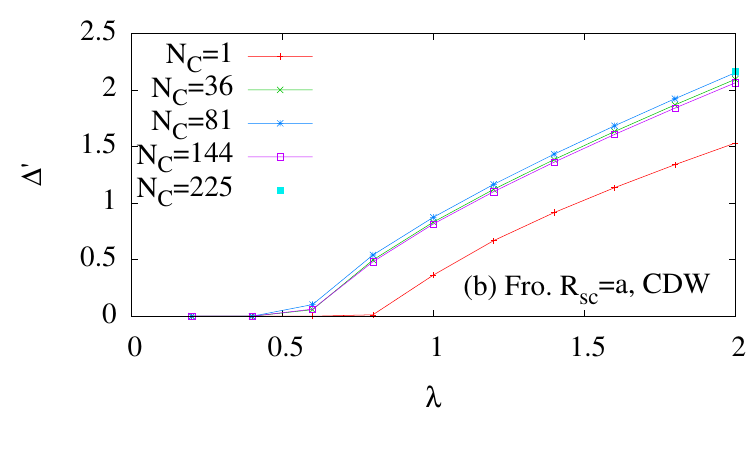}
\caption{(color online) (a) Spontaneous symmetry breaking (gap
  generation) in a graphene monolayer. The $y$-axis shows the
  spontaneously formed gap, $\Delta'$ and the $x$-axis the
  electron-phonon coupling $\lambda$. Here $T=0.02t$, $\Omega=0.01t$,
  $N_C=1$ and $\Delta=0$. (b) Effect of cluster size on the CDW state
  formed from a Fr\"ohlich interaction with $R_{sc}=a$. N.B. A single
  point is calculated for cluster size of $N_{C}=225$ to confirm
  convergence.}
\label{fig:spontaneous}
\end{figure}

The aim of the work presented here is to use dynamical cluster
approximation (DCA) to examine how charge density wave (CDW) gaps in
graphitic thin films vary with electron-phonon coupling. I take
$T=0.02t$, $\Delta=0.1t$ and $\Omega=0.01t$. Noting that $t$ is
typically on the order of an eV, these values correspond approximately
to room temperature, phonon frequencies, $\Omega$, of 10s of meV and
$\Delta$ a few hundred meV, consistent with thin films of materials
such as InSb, or reported gaps in some graphene on substrate systems. In the following, all results are for half filling.

I start by computing self-energies to show the relative contributions
of on- and off-site terms, and the effects of varying the interaction
range.  The computed self energies, including the relative
contributions of Hartree and Fock diagrams for $\lambda=2$ and $N_C=9$
resulting from a Holstein interaction are shown in
Fig. \ref{fig:cmpse}. The top two rows show real and imaginary parts
of the on-diagonal self energy, and the bottom two rows show the
off-diagonal self energy.  The real part of the on-diagonal Hartree
diagram is momentum and energy independent. It is the largest
magnitude element of the self energy matrix (contribution around
$2t$), and as it is frequency independent it directly contributes to
the enhancement to the gap by changing the effective local potential
on each sub-lattice, and it is also the main contributor to
spontaneous CDW order. The imaginary part of the Hartree diagram and
the off diagonal contributions are necessarily zero for the Holstein
interaction. The imaginary part of the Fock term (contribution $\sim
0.1t$) is still the most important contribution for some
properties. For example the inverse mass (not considered here) depends
on derivatives of the self energy, so this property will be given by
the Fock term. The Fock term is also the largest contribution to the
off-diagonal self energy with a contribution of around $0.01t$. Note
that when $\lambda_{A}=\lambda_{B}$, the self-energies have the
following symmetries: $\Sigma_{BA} = \Sigma_{AB}^{*}$ and
$\Sigma_{BB}=-\Sigma_{AA}^{*}$, so $\Sigma_{BA}$ and $\Sigma_{BB}$ are
not shown.

Fig. \ref{fig:cmpsefroh} is as Fig. \ref{fig:cmpse} for the longer
ranged Fr\"ohlich interaction (for $N_{C}=9$, the Hartree contribution
is $0.15t$). The real parts of the on-site self energies are
significantly smaller for the long-range Fr\"ohlich interaction than
for the Holstein model. Other differences are that the Fock
contribution to the off-site self energy is very small for the
Holstein interaction (N.B. it is not zero because of effective
off-site interactions mediated through the phonon self-energy),
whereas the off-site Fock self energy is of similar magnitude to (but
smaller than) the onsite Fock self-energy for the Fr\"ohlich
interaction ($\sim 0.04t$). The relative size of the on-site Fock
contribution drops off significantly relative to the Hartree diagram
as cluster size is increased, only contributing $\sim 4\%$ of the
total self energy at zero Matsubara frequency for $N_{C}=144$.

The main aim of this paper is to understand the role of
electron-phonon coupling range in the enhancement of gaps. Figure
\ref{fig:gapenhancecombined} shows how the enhancement varies with
interaction range and with cluster size. The gap size is calculated
directly from the value of the Hartree diagram, which causes a local
on-site potential energy shift, such that the gap,
$\Delta'\approx\Delta+\Sigma^{(H)}$, whereas the Fock term that
contributes the imaginary on-site self-energy changes the quasi-particle
lifetime. I have used Pad\'{e} approximants \cite{vidberg} to test for
any further gap contribution from the Fock term, which is very small
for the Holstein interaction, and for the Fr\"ohlich interaction
reduces from around 15\% for a cluster of $N_{C}=9$ to around 2\% for
cluster size $N_{C}=144$. Figure \ref{fig:gapenhancecombined}(a) shows
gap enhancement for a Holstein interaction with $\Delta/t=0.1$, for a
range of cluster sizes. N.B. The enhancements will be smaller for
systems with larger ionicity ($\Delta/t$)
\cite{hague2011b,hague2012a}. There are very small corrections due to
momentum dependence. As the screening radius, $R_{sc}$ increases, the
enhancement decreases. Increase in cluster size has no effect on the
gap in this set of diagrams for the Holstein
interaction. Fig. \ref{fig:gapenhancecombined}(b) shows results when
$R_{sc}=2a$. Essentially, the effective $\lambda$ (that goes like
$\lambda_{AA}(0)-\lambda_{AB}(0)$ in the Hartree diagram) decreases
with $R_{sc}$. The initial increase in the gap as $N_{C}$ is increased
is a result of the inhomogeneity in the effective electron-phonon
coupling across the Brillouin zone, which means that the value of
coupling is largest at the $\Gamma$ point, where it contributes most
to the Hartree diagram. For small clusters, averaging the coupling
across the Brillouin zone means that the coupling is under-estimated
at the zone center and over-estimated at the K point (see
Fig. \ref{fig:lambdazero}). Fig. \ref{fig:gapenhancecombined}(c) shows
gap enhancement for the long range Fr\"ohlich interaction
($R_{sc}\rightarrow\infty$). Even for the long range interaction, the
enhancement effects are significant. I note that the DMFT results
($N_{C}=1$) consistently underestimate the gap enhancement.

The phonon self-energy plays an important role in the gap
enhancement. Fig \ref{fig:gapenhancecombined}(a) also shows the
enhancement effect for the Holstein interaction when $N_C=1$ if the
polarization bubble (PB) is neglected and the curve can be compared
with the full theory with $N_C=1$. The phonon self energy augments the
enhancement by increasing the value of the phonon propagator at the
Brillouin zone center. Thus, increasing the electron self-energy,
which is proportional to the phonon propagator.

To show how the phonon self-energy varies within the Brillouin zone,
it is plotted in Fig. \ref{fig:phonse}. The variation is relatively
small, which indicates that the position space variation occurs over a
very small number of lattice sites, i.e. that only small clusters are
needed to capture the spatial variation of $\Pi$. To show how the
effective coupling is renormalised and depends on cluster size,
$D(\Qvec,0)$ is plotted in Figs. \ref{fig:effcouple} and
\ref{fig:effcouplefro}. The momentum dependence of the effective
coupling when the Holstein interaction is used is very weak
(Fig. \ref{fig:effcouple}). On the other hand, the effective coupling
is strongly momentum dependent for the Fr\"ohlich interaction. This
demonstrates that it is the effective coupling causes the spatial
variations that require large clusters to represent.

All results shown up to this point are calculated with electrons in
the monolayer coupling to phonons in the surface layer of ions in the
substrate only. Fig. \ref{fig:bulkcorr} shows corrections taking
account the finite depth of the bulk of the substrate rather than just
surface ions. The plot shows the effect of including interaction with
vibrations of surface ions only ($N_{z}=1$), and interaction with
vibrations in 3 layers ($N_{z}=3$) and 6 layers ($N_{z}=6$) of the
bulk of the substrate respectively (note that electrons still hop in
the monolayer, and do not hop into the substrate). The effect of the
bulk is to reduce the enhanced gap by around $20\%$.

Since the atoms on A and B sites may be different, their
electron-phonon interactions may also be different. I test the effect
of taking $\lambda_{A}\neq\lambda_{B}$, which is shown in
Fig. \ref{fig:lambdaneq}. In the figure, $\lambda_{A}$ is kept fixed,
while $\lambda_{B}$ is varied. The asymmetry induced between A and B sites
by the interaction means that self-energies are not symmetric between A and B sites as before, so
the chemical potential is varied iteratively during self consistency
to maintain half filling. Curves computed for these parameters are
compared with the enhancement when $\lambda_{A}=\lambda_{B}$, and the
average value of $\lambda$ is plotted on the $x$-axis to make the
comparison meaningful. As $\lambda_{B}$ is decreased, there is
initially a small reduction in the enhancement of around 20\%,
followed by an increase as $\lambda_{B}$ approaches zero. For a
comparable average $\lambda$, the enhancement is generally bigger than
for the case when the two couplings are the same. The reason why the
enhancement remains high is that large differences between the two
couplings significantly reduce the coupling between A and B sites
(which goes as $\sqrt{\lambda_{A}\lambda_{B}}$), and it is this
coupling that acts to reduce the enhancement in the Hartree term.

While this paper is primarily concerned with the modification of gaps
in graphene like (graphitic) materials with inherent ionicity, such as
thin films of III-V semiconductors, it is also interesting to
determine if gaps can spontaneously form from the electron-phonon
interaction. This is explored in Fig. \ref{fig:spontaneous}, which
shows spontaneous charge density wave (CDW) symmetry breaking. Panel
(a) shows DMFT results. For sufficient $\lambda$, a CDW state can be
found for all values of $R_{sc}$ (this is not visible in the figure
for large values of $R_{sc}$ because $\lambda$ values are too
small). The result shows that the full system with $\Delta\neq 0$ is
on the cusp of a CDW state, and this is why the gap is strongly
sensitive to the electron-phonon coupling. Panel (b) shows results for
$R_{sc}=1$ as the cluster size is increased. It may initially be of
surprise that a CDW state can be supported at finite temperature,
since Mermin-Wagner-Hohenberg (MWH) theorem does not allow for two
dimensional antiferromagnetism in Heisenberg models or 2D
superconductivity. However, detailed quantum Monte Carlo calculations
have shown that CDW order can be formed by the Holstein interaction at
half-filling on square lattices at finite temperature
\cite{noack1991a,vekic1992a,niyaz1993a}. In this case, MWH theorem
does not apply because the symmetry is discrete (i.e. the local charge
density at a specific time is determined by the number of electrons
and may be 0, 1 or 2) \cite{nowadnick2012a}.


To end this section, I note that as a minimum theory, it may be
sufficient to compute only the Hartree diagram (which dominates the
perturbation expansion) and the lowest order contribution to the
phonon self energy, $\Pi$, at $\Qvec=0$ and $\omega_s=0$ only (since only the
zero momentum Matsubara frequency and the phonon propagator
contributes to the Hartree diagram). However, there would still need to be
iteration over these diagrams to acheive self-consistency.





\section{Summary and conclusions}
\label{sec:conclusions}

In summary, I have investigated gap formation and enhancement in a
model of atomically thin graphitic materials. Electron-phonon coupling
and range has been varied, and the effects of higher order corrections
to the phonon propagator have been considered. The effect of
reintroducing fluctuations around the mean field limit has also been
investigated using the dynamical cluster approximation. Higher order
corrections to the perturbation theory increase the gap enhancement.
It is found that gaps are enhanced by electron-phonon interactions for
all interaction ranges, with the enhancement decreasing as interaction
range increases.

One of the driving factors of this enhancement is the proximity to a
charge density wave state for a material without ionicity ($\Delta=0$)
such as graphene. I have shown that sufficiently large coupling
between electrons and phonons can lead to spontaneous CDW order. This
instability to order shows why there are significant gap enhancements
at large coupling when ionicity is introduced. The existance of CDW
order at finite temperature in a 2D material such as graphene is
consistent with detailed quantum Monte Carlo results for a square
lattice \cite{noack1991a,vekic1992a,niyaz1993a,nowadnick2012a} and
this could be stabilized further at room temperature with small
interplane hopping of order 50meV (or around 1\% of the in-plane
hopping). Owing to the spontaneous symmetry breaking, an appropriately
layered heterostructure of graphene and a wide gap insulating material
such as BN might generate small spontaneous gaps of
useful size due to CDW formation.

In experiments, the strength of the electron-phonon coupling could be
varied in two ways. The first most obvious way to modify the coupling
between substrate and film is to change the substrate. Highly ionic
polarizable substrates would couple most strongly with the film,
leading to the strongest effects. While the distance between graphene
and substrate is of the order of 3\AA, the force between free
electrons and ions in a substrate (leading directly to electron-phonon
coupling) would be large. In fact, dimensionless electron-phonon
couplings of up to $\lambda=1$ have been reported in graphene on
substrate systems from angle resolved photoemission spectroscopy
studies (see Fig. 3 of Ref. \onlinecite{siegel2012a} and references
therein, note that much smaller interactions can be found with metal
substrates where polarizability is low and coupling with the substrate
is weak). An alternative way to dynamically decrease the
electron-phonon interaction range and increase coupling strength would
be to apply pressure to the film to move it closer to the substrate,
which could be simpler to achieve experimentally than growing films on
many different substrates.

I briefly mention that interactions with the vibrations of hydrogen
(and other) atoms that are used to functionalize graphene to make
graphane (and related materials) would be Holstein like, so part of
the gap in those materials may be phonon driven. This might be
testable by changing the isotope of the functionalizing atoms.

\begin{figure}
\includegraphics[width=85mm]{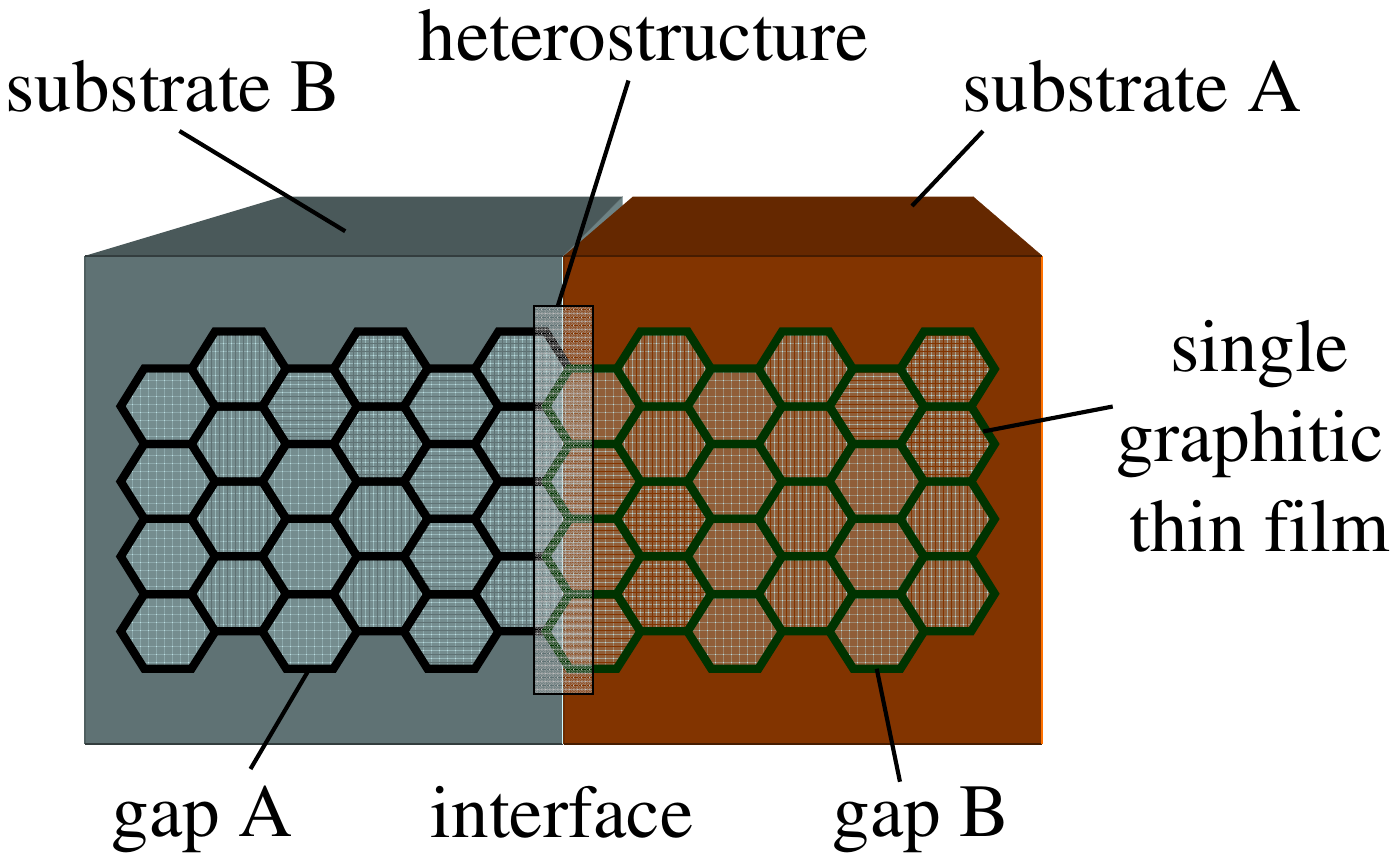}
\caption{Schematic showing the possible use of two substrates with different electron-phonon coupling, $\lambda$ to thin graphitic films to make a heterostructure in a single thin film of a graphitic material. This could be manufactured by laying down an interface between the two substrate materials before cleaving perpendicular to the interface and then depositing the thin film. The film above the substrate with the largest $\lambda$ would have a bigger gap, leading to a heterostructure within the interface region.}
\label{fig:heterostructure}
\end{figure}

The results here suggest that an interesting possibility would be to
use the electron-phonon interaction to make position dependent changes
to the bandstructure of the thin film (for example by adding a
spatially dependent superstrate with phonons that strongly couple to
electrons in the thin film), a method that is potentially easier to
control than trying to deposit neighboring thin films with interfaces
in the plane. Only tiny gap enhancements of around 20\% would be
needed so that proportional gap enhancements from the predictions made
here are similar to the proportional difference between gaps in GaAs
and AlGaAs \cite{ando1982a}, so it is plausible that thin film
heterostructures or quantum dots could be built up in this way (see
Fig. \ref{fig:heterostructure}). Another possibility would be to tune
inherent gaps in III-V semiconductors with the electron-phonon
interaction, so that they become optimal for applications such as
solar cells where the efficiency is highly sensitive to the gap
size. Clearly graphitic thin films warrant further study to assess
their full capability for novel electronics.

\section*{Acknowledgments}

I am pleased to acknowledge EPSRC grant EP/H015655/1 for funding and
useful discussions with Anthony Davenport, John Bolton and Adelina
Ilie.

\bibliographystyle{unsrt}
\bibliography{graphene_bn_dca}

\appendix

\end{document}